\documentclass[3p,times]{elsarticle}

\usepackage{amssymb}
\usepackage{amsthm}
\usepackage{amsmath}

\usepackage{color}

\usepackage[figuresright]{rotating}
\usepackage{amssymb}
\newfont{\fp}{msbm10 at 11pt}

\usepackage[T1]{fontenc}
\usepackage{amsfonts}

\usepackage{color, colortbl, framed}

\begin{document}

\begin{frontmatter}



\title{Codifference as a practical tool to measure interdependence}


\author[label1]{Agnieszka Wy{\l}oma{\'n}ska}
\author[label2]{ Aleksei Chechkin}
\author[label1]{Janusz Gajda}
\author[label3]{ Igor M. Sokolov}

\address[label1]{Hugo Steinhaus Center, Institute of Mathematics and Computer Science,\\     
    Wroclaw University of Technology, Wroclaw, Poland \\
    agnieszka.wylomanska@pwr.wroc.pl\\
    janusz.gajda@pwr.wroc.pl\\
   }
	\address[label2]{ Akhiezer Institute for Theoretical Physics, National Science Center "Kharkov Institute of Physics and Technology"\\
Kharkov 61108, Ukraine, and\\ Max Planck Institute for Physics of Complex Systems, Noethnitzer Str 38, D-01187 Dresden, Germany\\
achechkin@kipt.kharkov.ua}
	\address[label3]{ Institut für Physik, Humboldt-Universität zu Berlin, Newtonstrasse 15, D-12489 Berlin, Germany\\   
igor.sokolov@physik.hu-berlin.de}
\begin{abstract}
Correlation and spectral analysis represent the standard tools to study interdependence 
in statistical data. However, for the stochastic processes with heavy-tailed distributions 
such that the variance diverges, these tools are inadequate. The heavy-tailed processes 
are ubiquitous in nature and finance.
We here discuss codifference as a convenient measure to study statistical interdependence, 
and we aim to give 
a short introductory review of its properties. By taking different known stochastic processes 
as generic examples, 
we present explicit formulas for their codifferences. We show that for the Gaussian processes 
codifference is
equivalent to covariance. For processes with finite variance these two measures behave similarly
with time. For the processes with infinite variance the covariance does not exist, however,
the codifference is relevant. We demonstrate the practical importance of the codifference
by extracting this function from simulated as well as from real experimental data. We conclude that 
the codifference serves as a convenient practical tool to study interdependence 
for stochastic processes 
with both infinite and finite variances as well.

\end{abstract}
\end{frontmatter}
\section{Introduction}

Stochastic processes with diverging variance are ubiquitous in nature and finance. A remarkable example 
is an alpha-stable L\'evy motion, or L\'evy flights, that is
the class of non-Gaussian Markovian random processes whose stationary independent increments 
are distributed according to L\'evy stable distributions \cite{levy}. L\'evy stable laws are important 
for three fundamental properties: (i) according to generalized central limit theorem, they form 
the basin of attraction for sums of random variables with diverging variance \cite{Gnedenko}; 
(ii) the probability density functions of L\'evy stable laws decay in asymptotic power-law form 
and thus appear naturally in the description of many fluctuation processes with largely 
scattering statistics characterized by bursts or large outliers; (iii) L\'evy flights 
are statistically self-affine, a property used for the description of random fractal 
processes. Examples of L\'evy flights range
from light propagation in fractal medium called L\'evy glass \cite{glass} and plasma 
fluctuations in fusion devices \cite{plasma1,plasma2,plasma3,plasma4} to circulation of dollar bills
\cite{dollar} and behavior of the marine verterbrates in response to patchy distribution of food
resources \cite{marine}; more examples can be found, e.g., in \cite{anotrans, encyclopedia}.  
The L\'evy flight dynamics can also stem from a simple Brownian random walk in systems whose
operational time typically grows superlinearly with physical time $t$ \cite{sok1, sok2}.

Other prominent examples of the processes with heavy-tailed distributions 
are L\'evy Ornstein-Uhlenbeck
(OU) process describing the overdamped harmonic oscillator driven by alpha-stable
L\'evy noise, and fractional L\'evy stable motion. The L\'evy OU process 
is a natural generalization of the Gaussian OU process; 
such generalization 
has gained popularity, e.g., in finance \cite{vas, barndorff, Ilya}. The weakly damped
harmonic oscillator driven by L\'evy noise was discussed in \cite{chego, soebdy}. The 
L\'evy OU process 
accounts for interdependence (association) of exponential type. On the contrary, fractional motions 
and fractional noises 
have an infinite span of
interdependence \cite{mandelbrot, samorodnitsky}. Fractional processes are also widely spread 
in applications \cite{jeon1, goychuk, jeon2}. Indeed, in a large class of many-particle 
systems whose 
overall dynamics is Markovian, the probe particle coupled with the rest of the system through space 
correlations exhibit fractional motion with long-ranged non-Markovian memory 
effects \cite{taloni1, taloni2, taloni3}. 
Fractional L\'evy stable motion with long-range dependence was detected 
in beat-to-beat heart rate 
fluctuations \cite{peng}, in solar flare time series \cite{solar}, and was shown to be
a model qualitatively mimicking self-organized criticality signatures in data \cite{nick}.  

What is the measure of interdependence for the processes with infinite variance? Apparently, correlation 
or spectral power analysis, strictly speaking, can not be used. The alternative measures of dependence are
rarely discussed in application-oriented literature. 

The notion of covariance (CV) used in correlation analysis, can be generalized
for the alpha-stable L\'evy process, leading to the notion of covariation
\cite{samorodnitsky,kokoszka1}. Its
definition is based on the L\'evy measure, and its practical usefulness
is limited. Some results related 
to the covariation of autoregressive process with L\'evy stable distribution 
are presented in \cite{nowicka_wylomanska}. 
The other measure is the L\'evy correlation cascade \cite{eliazar}. It is defined for 
infinitely divisible processes, 
and the properties of that measure are discussed in \cite{mag1} in detail, see 
also \cite{wylomanska_acta}. 
However, similar to covariation, the L\'evy correlation cascade is of limited practical 
value because 
of complicated definition based on the L\'evy measure of the underlying process. 

Our paper deals with another measure of interdependence called codifference (CD). It is based 
on the characteristic function of a given process, therefore it can be used not only for alpha-stable 
processes. Moreover, the codifference in the Gaussian case reduces to the classical covariance, 
so it can be treated as the natural extension of the well-known measure. On the other hand, according 
to the definition, it is easy to evaluate the empirical codifference which is based on the empirical 
characteristic function of the analyzed data. It is worth to mention that the codifference is closely 
related to the so-called dynamical functional used to study ergodic properties of stochastic 
processes \cite{wer1,wer2,samorodnitsky}. In our paper we aim to give an introduction to the concept 
of codifference and to show its usefulness for analyzing interdependence not only for the processes 
with infinite variance but for those with finite variance, as well.

The rest of the paper is structured as follows: In Section 2 we give definitions of codifference. 
In Section 3 we compare autocovariance and autocodifference for
Gaussian processes and for non-Gaussian processes with finite variance. In Section 4 we present 
the autocodifference
for processes with infinite variance. The method how to estimate autocodifference from experimental 
data is described in Section 5 together with the examples taken from simulations. In Section 6
we present the results of real data analysis for two processes with infinite variance. The conclusions
are presented in Section 7, and several technical details of calculating autocodifference are
collected in the Appendix.

%\begin{figure}
%\begin{center}
%\includegraphics[width=0.7\textwidth]{figure1j.eps}
%\caption{Comparison of MSD obtained from Monte Carlo simulation of the process $X(t)$ with the theoretical results in \eqref{meansquaredisplacementcalculated}. One can observe that MSD of the process $X(t)$ behaves as in subdiffusive case. Parameters are: $\alpha=1.8, \lambda=0.5,\gamma=0.85$.}
%\label{figure1j}  
%\end{center}
%\end{figure}
\section{Definitions}
We start from the definition of codifference for the symmetric alpha-stable (S$\alpha$S) random variables. Let us remind 
that the random variable $X$ which is S$\alpha$S with parameter $0<\alpha\leq 2$ and scale parameter $\sigma_{X}>0$ 
has the following characteristic function $\Phi_{X}(k)$ \cite{Gnedenko}:
\begin{eqnarray}
\Phi_{X}(k)=<\exp(ikX)>=\exp\{-\sigma_{X}^{\alpha}|k|^{\alpha}\},
\end{eqnarray}
where $<...>$ denote ensemble average or average over realizations. 
The codifference 
of two jointly S$\alpha$S, $0<\alpha\leq 2$, random variables $X$ and $Y$ is defined as follows \cite{samorodnitsky}:
\begin{equation}
\label{cod1}
CD(X,Y)=\sigma_{X-Y}^{\alpha}-\sigma_{X}^{\alpha}-\sigma_{Y}^{\alpha},
\end{equation}
where $\sigma_{X},\sigma_{Y}$ and $\sigma_{X-Y}$ denote, respectively the scale parameters 
of $X,Y$ and $X-Y$.
The codifference  can be also  defined in the language of the characteristic function \cite{kokoszka1,nowicka_wylomanska,mag1}
\begin{equation}
\label{Codifference}
CD(X,Y)=\ln(<exp\{iX\}>)+\ln(<exp\{-iY\}>)-\ln(<exp\{i(X-Y)\}>).
\end{equation}
Thus, the definition given in Eq.(\ref{cod1}) can be extended to a more general class of random variables, and in the further analysis we use the representation given in (\ref{Codifference}).
 We note, that  
one of the reasons to take the minus sign, i.e., $-iY$ in the second term
of the right-hand side of Eq.(\ref{Codifference}) is to ensure that CD is reduced to covariance in the Gaussian case
(with the minus sign); see below.\\
It is worth to mention that the codifference shares useful properties \cite{samorodnitsky}
\begin{itemize}
\item it is always well-defined, since the definition of $CD(X,Y)$ is based on the characteristic functions 
of appropriate random variables;
\item if the random variables are symmetric, then $CD(X,Y)=CD(Y,X)$;
\item if $X$ and $Y$ are independent and jointly S$\alpha$S, then $CD(X,Y)=0$. Conversely, if $CD(X,Y)=0$ 
and $0<\alpha<1$, then $X$ and $Y$ are independent. When $1<\alpha<2$, $CD(X,Y)=0$
does not imply that $X$ and $Y$ are independent. We note that in case of two Gaussian random variables 
(i.e. S$\alpha$S with $\alpha=2$) the independence of $X$ and $Y$ is equivalent to zero covariance. 
\end{itemize}
The above properties confirm that the codifference is an appropriate mathematical tool for measuring 
the dependence between alpha-stable  random variables as well as random variables from more 
general class of distributions (e.g., infinitely divisible).  In the literature one can also find the 
generalized codifference which is defined as \cite{samorodnitsky}:
\begin{equation}\label{generalized}GCD(X,Y;\theta_1,\theta_2)=
\ln(<exp\{i\theta_1X\}>)+\ln(<exp\{i\theta_2Y\}>)-\ln(<exp\{i(\theta_1X+\theta_2Y)\}>),\end{equation}
where $\theta_1,\theta_2\in R$. The generalized codifference is useful in cases when the random 
variables 
considered have different scales or different units \cite{soebdy}. Thus, the proper choice 
of $\theta_1$ and $\theta_2$ parameters 
allows one to analyze the interdependence at the same scale. Also, to make the measure of
dependence invariant against the change of units, it is reasonable to take $\theta_1$ equal
to $\sigma_X^{-1}$ and $\theta_2$ equal to $\sigma_Y^{-1}$. Equation (\ref{generalized})
will then define the "cosum" of non-dimensional, normalized variables \cite{soebdy}.
The usage of "cosum" has an additional advantage when e.g., asymmetric alpha-stable L\'evy
variables are discussed, since the sum of two such variables has a distribution
which differ only by a scaling factor, whereas the difference has a distribution which
belongs to a different class. In the present paper we however stick to the definition 
(\ref{Codifference}).\\
For a stochastic process $\{X(t)\}$, the measure $CD(X(t),X(s))$ called autocodifference is defined as \footnote{In this paper we denote a stochastic 
process as $\{X(t)\}$ and the value of the process at time $t$ as $X(t)$.}
\begin{equation}
\label{AutoCodifference}
CD(X(t),X(s))=\ln(<exp\{iX(t)\}>)+\ln(<exp\{-iX(s)\}>)-\ln(<exp\{i(X(t)-X(s))\}>).
\end{equation}
For the stationary process the autocodifference $CD(X(t),X(s))$ depends on $|t-s|$.

\section{Autocodifference versus autocovariance for processes with finite second moment}
In this Section we consider autocodifference for the several processes with finite variance,
which are widely used in applications. We restrict ourselves with stationary processes and processes
with stationary increments.
\subsection{Gaussian processes}
For the Gaussian processes the autocodifference is simply reduced to autocovariance with the minus sign. Indeed, let $\{X(t)\}$ 
be a Gaussian process with mean $f(t)$ and variance $g(t)$, then the characteristic function reads:
\begin{eqnarray}\label{char}\Phi_{X(t)}(k)=<e^{ikX(t)}>=e^{ikf(t)-g(t)k^2/2}.\end{eqnarray}
Then, for fixed $t$ and $s$ ($s<t$) the increments $X(t)-X(s)$ have also Gaussian distribution with the mean $f(t)-f(s)$ and variance equal to $g(t)+g(s)-2cov(Y(t),Y(s))$. 
Therefore, from  Eq.(\ref{AutoCodifference}) we get:
\begin{eqnarray*}
CD(X(t),X(s))&=&+if(t)-\frac{g(t)}{2}-if(s)-\frac{g(s)}{2}\\
&-&i(f(t)-f(s))+\frac{g(t)+g(s)-2cov(X(t),X(s))}{2}=-cov(X(t),X(s)).\end{eqnarray*}
Below we give some examples of the autocodifference for Gaussian processes.
\subsubsection  {Gaussian white noise process}
The process $\{b(t)\}$ is called white Gaussian noise  if for each $t$ the random variable $b(t)$ 
has Gaussian distribution with zero mean and finite variance $\sigma^2$, and for each $t\neq s$ 
the values $b(t)$ and $b(s)$ are uncorrelated. Thus, the autocodifference is given by:
\begin{eqnarray}
CD(b(t),b(s))=-cov(b(t),b(s))=\left\{\begin{array}{lll}  -\sigma^2, & \mbox{if} & t=s,\\
0, &  &\mbox{otherwise.}
\end{array}\right.
\end{eqnarray}

\subsubsection{Ordinary Brownian motion}
The process is called ordinary Brownian motion if it has stationary independent increments possessing Gaussian distribution. The standard ordinary Brownian motion $\{B(t)\}$ has zero mean and variance equal to $t$. In this case we have:
\begin{eqnarray}\label{brownianmotion}CD(B(t),B(s))= -cov(B(t),B(s))=-\min\{t,s\}.\end{eqnarray}
\subsubsection{Gaussian Ornstein-Uhlenbeck process}
The Gaussian Ornstein-Uhlenbeck process is defined as a stationary solution of the Langevin equation for
the overdamped oscillator:
\begin{eqnarray}\frac{dY(t)}{dt}+\lambda Y(t)=b(t),~~\lambda>0\end{eqnarray}
where $\{b(t)\}$ is the Gaussian white noise (heuristically $b(t)=dB(t)/dt$). The Gaussian Ornstein-Uhlenbeck process has the following moving average
representation:
\begin{eqnarray}Y(t)=\int_{-\infty}^te^{-\lambda(t-s)}d\tilde{B}(s),\end{eqnarray}
where $\tilde{B}(t)$ is an extension of the Brownian motion for the negative axis, that is
\begin{eqnarray}
\tilde{B}(t)=\left\{\begin{array}{lll} B(t), & \mbox{when} & t\geq 0,\\
B(-t), &  &\mbox{otherwise.}
\end{array}\right.
\end{eqnarray}
We remind that $\{Y(t)\}$ is the only
stationary and Markovian - Gaussian stochastic process, according to the Doob's theorem \cite{dob}. The autocodifference is given by:
\begin{eqnarray}\label{oug}CD(Y(t),Y(s))=-cov(Y(t),Y(s))=-\frac{e^{-\lambda|t-s|}}{2\lambda}.\end{eqnarray}
 
\subsubsection{Fractional Brownian motion }
The fractional Brownian motion is a zero mean Gaussian process $\{B_H(t)\}$ defined as follows \cite{mandelbrot}:
\begin{eqnarray}\label{fbm}
B_H(t)=\int_{-\infty}^{0}\left\{(t-u)^{H-1/2}-(-u)^{H-1/2}\right\}dB(u)+\int_{0}^t(t-u)^{H-1/2}dB(u),
\end{eqnarray}
where $\{B(t)\}$ is the classical Brownian motion. The $H$ parameter called Hurst exponent  is a real number from the interval  $(0, 1)$. 
The process is self-similar and reduces to ordinary Brownian motion, if $H=1/2$. The parameter $H$ controls the type
of diffusion, namely the process is superdiffusive for $H>1/2$ while subdiffusive for $H<1/2$.  
The autocodifference takes the following form:
\begin{eqnarray}CD(B_H(t),B_H(s))=-cov(B_H(t),B_H(s))=-\frac{k(H)}{2}(|t|^{2H}+|s|^{2H}-|t-s|^{2H}),\end{eqnarray}
where $k(H)=Var(B_{H}(1))=\int_{0}^{\infty}\left|(1+x)^{H-1/2}-x^{H-1/2}\right|^2dx+\frac{1}{2H}$.

\subsubsection{Fractional Gaussian noise}
The fractional Gaussian noise $\{b_H(t)\}$ is defined heuristically as the derivative of the fractional Brownian motion, 
i.e. $b_{H}(t)=dB_H(t)/dt$. Therefore, the autocodifference of the process takes the following form \cite{marek_ja}:
\begin{eqnarray}\label{fgn}CD(b_H(t),b_H(s))=-cov(b_H(t),b_H(s))=\frac{k(H)}{2}(|t-s+1|^{2H}-2|t-s|^{2H}+|t-s-1|^{2H}),\end{eqnarray}
where $k(H)=Var(b_{H}(t))=\int_{0}^{\infty}\left|(1+x)^{H-1/2}-x^{H-1/2}\right|^2dx+\frac{1}{2H}$. Detailed derivation 
of covariance structure for both fractional Brownian motion and fractional Gaussian noise is presented 
in \cite{marek_ja, bib:Oksendal}.\\

\subsection{Non-Gaussian processes}
\subsubsection{Poisson process}
The Poisson process with intensity $\lambda$ is a continuous-time counting process $\{P(t)\}$ which has stationary, 
independent increments \cite{bib:Kingman, stat_tools}. The increment $P(t)-P(s)$ (for $t>s$) has Poisson distribution 
with parameter $\lambda(t-s)$. The covariance function of the Poisson process takes the following form: 
\begin{eqnarray}cov(P(t),P(s))=\min\{t,s\}\lambda.\end{eqnarray}
The characteristic function of the Poisson process is given by $\Phi_{P(t)}(k)=e^{t\lambda(e^{ik}-1)}$, 
therefore the autocodifference takes the following form:
\begin{eqnarray}CD(P(t),P(s))=-2\lambda \min\{t,s\}(1-\cos 1).\end{eqnarray}
Some interesting extensions of Poisson process one can find in \cite{me2}.
\subsubsection{Tempered stable L\'evy process}
The tempered stable distributions were suggested in \cite{bib:Mantegna} 
and studied in \cite{bib:Koponen} in more detail. The general mathematical description of this class 
of processes was established in \cite{rosinski}. In our paper we consider the tempered stable distribution 
with the  L\'evy triplet $(\kappa^2, \nu, \gamma)$, which is defined as follows \cite{kim}:\\
\begin{eqnarray*}
\kappa=0\\
\nu(dx)=(\tilde{C}_{+}e^{-\lambda_{+}x}1_{x>0}+\tilde{C}_{-}e^{-\lambda_{-}|x|}1_{x<0})\frac{dx}{|x|^{\alpha+1}}\\
\gamma=m-\int_{|x|>1}x\nu(dx),
\end{eqnarray*}
where $\tilde{C}_{+},\tilde{C}_{-},\lambda_{+},\lambda_{-}>0$, $\alpha\in(0,2)$ and $m\in R$. \\
A L\'evy process (i.e. process with independent stationary increments) having tempered stable distribution
 is called a tempered stable process with parameters $\alpha,\tilde{C}_{+},\tilde{C}_{-},\lambda_{+},\lambda_{-},m$.

For simplicity we consider the special case, namely we assume here $m=0$ and $\lambda_{+}=\lambda_{-}=\lambda$, $\alpha> 1$. Moreover if we substitute  

$$C_{+}=\frac{1}{\Gamma(-\alpha)}\tilde{C}_{+},~~C_{-}=\frac{1}{\Gamma(-\alpha)}\tilde{C}_{-},$$ then it can be shown that in this case the characteristic function $\phi_{T(t)}(k)=Ee^{ikT(t)}$ of the tempered stable process $\{T(t)\} $ takes the following form:\\

\begin{eqnarray}\label{fourier}
\phi_{T(t)}(k)=\exp\left\{t\left[i\alpha k\lambda^{\alpha-1}(C_{+}-C_{-})+C_{+}(\lambda-ik)^{\alpha}+C_{-}(\lambda+ik)^{\alpha}-\lambda^{\alpha}(C_{+}+C_{-})\right]\right\}.
\end{eqnarray}
The tempered stable L\'evy process (called also truncated L\'evy flight) is a process $\{T(t)\}$ with independent stationary increments 
possessing tempered stable distribution described above. As an example, we  analyze the symmetric tempered stable process $\{T(t)\}$,
i.e. such that the characteristic function of $T(t)$ is given by:
\begin{eqnarray}\label{char_example}\Phi_{T(t)}(k)=\exp\left\{t\left[(\lambda+ik)^{\alpha}+(\lambda-ik)^{\alpha}-2\lambda^{\alpha}\right]\right\}.
\end{eqnarray}
The tempered stable process is an extension of the alpha-stable L\'evy process, and for $\lambda=0$ the random 
variable  $T_{\alpha,\lambda}(t)$ reduces to the appropriate alpha-stable random variable, see \cite{meer}.\\
The autocovariance of the analyzed tempered stable process has the following form:
\begin{eqnarray}cov(T(t),T(s))=2\alpha(\alpha-1)\lambda^{\alpha-2}\min\{s,t\}.\end{eqnarray}
The autocodifference is given by
\begin{eqnarray}\label{tscd}CD(T(t),T(s))=2[(\lambda-i)^{\alpha}+(\lambda+i)^{\alpha}-2\lambda^{\alpha}]\min\{t,s\}.\end{eqnarray}
For more details see Appendix, Section A1.
\subsubsection{Laplace motion}
The Laplace motion (called also variance gamma process) is a process $\{\Lambda(t)\}$, $t\geq0$ defined as follows \cite{lap}:
\begin{equation}
\label{LaplaceMotion}
\Lambda(t)=B(G(t)),
\end{equation}
where $\{B(\cdot)\}$ is an ordinary Brownian motion, and $\{G(t)\}$ is a Gamma process with parameters $\gamma, \lambda$ defined 
as a pure jump L\'evy increasing process with increments having Gamma distribution. It has the following characteristic function:
\begin{equation}
\Phi_{G(t)}(k)=<\exp(ikG(t))>=(1-ik/\lambda)^{-\gamma t}, k\in R.
\end{equation}
The Laplace motion has been successfully applied in the modeling of credit risk in structural models \cite{vg,vg2,vg1}. 
Some extensions of this process one can find, e.g., in \cite{me1}.\\
The process $\{\Lambda(t)\}$ has a zero mean and stationary independent increments. The covariance 
function reads
\begin{eqnarray}
cov(\Lambda(t),\Lambda(s))=\frac{\Gamma(\gamma s+1)}{\lambda\Gamma(\gamma s)}, \mbox{$s<t$},\end{eqnarray}
where $\Gamma(a)$ is a gamma function, %Because
%\begin{eqnarray}
%\ln <\exp(i(\Lambda(t)-\Lambda(s))>&=&\ln <\exp(i(\Lambda(t-s))>=\ln \exp(-1/2\Gamma(t-s)))\\\nonumber
%&=& (-\gamma(t-s))\ln (1+1/(2\lambda))\end{eqnarray}
while the autocodifference reads
\begin{eqnarray}
CD(\Lambda(t),\Lambda(s))=-2\gamma s\ln (1+1/(2\lambda)), \mbox{$s<t$}.
\end{eqnarray}
See Appendix, Section A2, for more details.

\section{Autocodifference for processes with infinite variance}
In this section we analyze the processes for which the autocovariance function is not defined, and then the 
autocodifference is the main measure of dependence.
\subsection{White L\'evy noise}
The white L\'evy noise $\{l_{\alpha}(t)\}$, $t\geq 0$, is a   process such that for each $s\neq t$ the random variables $l_{\alpha}(s)$ and $l_{\alpha}(t)$ are independent, and for each $t$ the random variable $l_{\alpha}(t)$ has the following   characteristic function:
\begin{eqnarray}
\Phi_{l_{\alpha}(t)}(k)=\left\{\begin{array}{lll} \exp\{-\sigma^{\alpha}|k|^{\alpha}(1-i\beta sgn(k)\tan\frac{\pi\alpha}{2})+ik\mu\} & \mbox{if} & \alpha\neq 1,\\
\exp\{-\sigma |k|(1-i\beta\frac{2}{\pi}sgn(k)\ln|k|)+ik\mu)\} &  \mbox{if} & \alpha=1,
\end{array}\right.
\end{eqnarray}
where $0<\alpha\leq 2$ is the stability index, $\sigma>0$ the scale parameter, $-1\leq\beta\leq 1$ the skewness parameter, 
and $\mu\in R$ is the shift parameter. \\
The autocodifference reads as
\begin{eqnarray}\label{cd_sn}
CD(l_{\alpha}(t),l_{\alpha}(s))=\left\{\begin{array}{lll} -2\sigma^{\alpha}, & \mbox{if} & t=s,\\
0, & & \mbox{otherwise.}
\end{array}\right.
\end{eqnarray}
It is worth to mention that the above formula is valid not only for symmetric L\'evy noise.
%-\left[-\sigma^{\alpha}(1-i\beta \tan\frac{\pi\alpha}{2})+i\mu\right]-\left[-\sigma^{\alpha}(1+i\beta \tan\frac{\pi\alpha}{2})+i\mu\right]\]
%
%
\subsection{L\'evy flights}
The L\'evy flight (called also alpha-stable L\'evy motion) is the process $\{L_{\alpha}(t)\}$, 
$t\geq 0$, with independent stationary increments possessing alpha-stable distribution, i.e. in general case for each $t$ the random variable $L_{\alpha}(t)$ has the stable distribution with index of stability $\alpha$, scale parameter $t^{1/\alpha}\sigma$, skewness $\beta$ and shift parameter $\mu =0$. 
The autocodifference for the analyzed process takes the form
\begin{eqnarray}\label{lf}CD(L_{\alpha}(t),L_{\alpha}(s))=-2\sigma^{\alpha}\min\{t,s\}.\end{eqnarray}
%-\sigma^{\alpha}(t-s)\left[(1-i\beta\tan\frac{\pi\alpha}{2})+i\mu\right]+\sigma^{\alpha}t\left[(1-i\beta )\tan\frac{\pi\alpha}{2})+i\mu\right]\]
%\[\sigma^{\alpha}s\left[(1-i\beta \tan\frac{\pi\alpha}{2})-ik\mu\right]\]
%
%
Let us note,  for $\alpha=2$ and $\sigma=1/\sqrt{2}$ (parameters of symmetric alpha-stable distribution corresponding 
to the standard Gaussian one) the above formula reduces to (\ref{brownianmotion}).

\subsection{L\'evy Ornstein-Uhlenbeck process}
The L\'evy Ornstein-Uhlenbeck process $\{Y_{\alpha}(t)\}$ is defined via the following Langevin equation:
\begin{eqnarray}\label{o-u-s}\frac{dY_{\alpha}(t)}{dt}+\lambda Y_{\alpha}(t)=l_{\alpha}(t),~~\lambda>0,\end{eqnarray}
where $\{l_{\alpha}(t)\}$ is white L\'evy noise. The process was examined, for example, in \cite{ou}.  
The stationary solution of equation (\ref{o-u-s}) reads
\begin{eqnarray}\label{rozw11}
Y_{\alpha}(t)=\int_{-\infty}^te^{-\lambda(t-u)}d\tilde{L}_{\alpha}(u),
\end{eqnarray}
where 
\begin{eqnarray}
\tilde{L}_{\alpha}(t)=\left\{\begin{array}{lll} L_{\alpha}(t), & \mbox{when} & t\geq 0,\\
L_{\alpha}(-t), &  &\mbox{otherwise.}
\end{array}\right.
\end{eqnarray}
For simplicity, we assume that in the considered case $\{L_{\alpha}(t)\}$ is symmetric alpha-stable L\'evy process i.e. process, 
for which the increments have alpha-stable distribution with the $\mu=0$, $\beta=0$ and scale parameter $\sigma$. 
Then the autocodifference takes the form
\begin{eqnarray}\label{cd_ou}CD(Y_{\alpha}(t),Y_{\alpha}(s))=-\sigma^{\alpha}\frac{1+e^{-\lambda\alpha |t-s|}-|1-e^{-\lambda|t-s|}|^{\alpha}}{\lambda \alpha}.\end{eqnarray}
For more details see Appendix, Section A3.
We note that for $\alpha=2$ and $\sigma=1/\sqrt{2}$ (parameters of symmetric alpha-stable distribution corresponding 
to the standard Gaussian one) the above formula reduces to (\ref{oug}).\\

\subsection{Fractional L\'evy motion}
The fractional L\'evy motion is a process $\{L_{\alpha,H}(t)\}$ defined for any $0<H<1$ as follows \cite{samorodnitsky,krzysiek,burn11,JW94}:
\begin{eqnarray}\label{FLSM}
L_{\alpha,H}(t)=\int_{-\infty}^{0}\left\{(t-u)^{H-1/\alpha}-(-u)^{H-1/\alpha}\right\}dL_\alpha(u)+\int_{0}^t(t-u)^{H-1/\alpha}dL_\alpha(u),
\end{eqnarray}
where $\{L_\alpha(t)\}$ is the symmetric alpha-stable L\'evy process with index of stability $\alpha\in(0,2)$. 
For simplicity we assume $\sigma=1$. The process $\{L_{\alpha,H}(t)\}$ is self-similar, stationary-increment process 
with infinite second moment. Similar to fractional Brownian motion, the parameter $H$ controls the diffusion law, namely
for $H<1/\alpha$ and $H>{1}/{\alpha}$ the fractional L\'evy motion exhibits sub- and superdiffusive behavior, respectively \cite{bur_si_wer}. 
The autocodifference for fractional L\'evy motion has the following form: %\cite{maej}:
%\begin{eqnarray}CD(L_{\alpha,H}(t),L_{\alpha,H}(0))\rightarrow k(\alpha,H)|t|^{\alpha(H-1)}, t\rightarrow\infty\end{eqnarray}
%where $k(\alpha,H)$ is an appropriate constant.
\begin{eqnarray}\label{flm}CD(L_{\alpha,H}(t),L_{\alpha,H}(s))=k(H,\alpha)\left(|t-s|^{\alpha H}-|t|^{\alpha H}-|s|^{\alpha H}\right),\end{eqnarray}
where $k(H,\alpha)=\int_{0}^{\infty}\left| (1+u)^{H-1/\alpha}-u^{H-1/\alpha}\right|^{\alpha}du+\frac{1}{H\alpha}$.\\
See Appendix, Section A4, for more details.\\
Let us remind that for $H=1/\alpha$ the fractional L\'evy motion reduces to L\'evy flights. In this case 
the autocodifference (\ref{flm}) takes the form $-2\min\{s,t\}$ which is consistent with formula (\ref{lf}) 
under the assumption $\sigma=1$.
%where $C_1,C_2$ are constants. \\
%
%
\subsection{Fractional L\'evy noise}
The fractional L\'evy noise $\{l_{\alpha,H}(t)\}$ is defined heuristically as the derivative of the fractional L\'evy motion $\{L_{\alpha,H}(t)\}$. 
For large $t$ the autocodifference has a power law form \cite{ast}:\\
If either $0<\alpha\leq 1$, $0<H<1$ or $1<\alpha<2$, $1-\frac{1}{\alpha(\alpha-1)}<H<1$, $H\neq1/\alpha$, then for $t\rightarrow \infty$
\begin{eqnarray}\label{asym1}CD(l_{\alpha,H}(t),l_{\alpha,H}(0))\sim C t^{\alpha H-\alpha}.\end{eqnarray}
If $1<\alpha<2$ and $0<H<1-\frac{1}{\alpha(\alpha-1)}$, then for $t\rightarrow \infty$
\begin{eqnarray}\label{fln}CD(l_{\alpha,H}(t),l_{\alpha,H}(0))\sim D t^{H-1/\alpha-1},\end{eqnarray}
where the constants $C$ and $D$ are time-independent.\\
Note that, for $\alpha=2$ the fractional L\'evy noise reduces to fractional Gaussian noise. In this case, the autocodifference 
given in (\ref{fgn}) behaves as $t^{2H-2}$ for $t\rightarrow \infty$ which is consistent with (\ref{asym1}) \cite{ast}. 
\subsection{Superdiffusion continuous time random walk-like process}
We consider superdiffusive continuous time random walk-like process which is defined as follows \cite{sok1}:
\begin{eqnarray}\label{ctrwlike}
Z(t)=B(L_{\alpha}(t)),
\end{eqnarray}
where 
$\{B(t)\}$ is the ordinary Brownian motion and $\{L_{\alpha}(t)\}$ the L\'evy flight process described in Section 4.2, 
with $0<\alpha<1$, $\beta=1$, $\mu=0$ and scale parameter $\sigma>0$, i.e. the process for which the Laplace transform is given by:
\begin{eqnarray}\label{lt}\mathcal{L}_{L_{\alpha}(t)}(k)=<\exp\{-kL_{\alpha}(t)\}>=\exp\{-t(\sigma|k|)^{\alpha}\}.\end{eqnarray}
The processes $\{B(t)\}$ and $\{L_{\alpha}(t)\}$ are independent.\\
The process $\{L_{\alpha}(t)\}$ in definition (\ref{ctrwlike}) plays the role of time (i.e. is should be non-negative increasing process), 
therefore the above restrictions on the parameters have to be assumed.

Since Brownian motion and L\'evy flight processes have independent stationary increments, then 
the process $\{Z(t)\}$ also possesses this property \cite{sato}. 
The possible applications of superdiffusion continuous time random walk-like process are transport 
in heterogeneous catalysis, micelle systems, reactions and transport in polymer systems 
under conformational motion and dynamical systems \cite{sok2}.

Since $<B^2(L_{\alpha}(t))>=<L_{\alpha}(t)>$ and the mean of $L_{\alpha}(t)$ does not exists, then 
the process $\{Z(t)\}$ has infinite second moment. 
The autocodifference for the process $\{Z(t)\}$ is given by:
\begin{eqnarray}CD(Z(t),Z(s))=-2^{-\alpha+1}\sigma^{\alpha}\min\{t,s\}.\end{eqnarray}
See Appendix, Section A5, for more details.
\section{How to estimate codifference from  data}
We define an estimator of autocodifference in the form:
\begin{eqnarray}
\label{CodifferenceEstimator}
\widehat{CD}(X(t),X(s))=\ln(\hat{\phi}(1,0, X(t), X(s)))+\ln(\hat{\phi}(0,-1, X(t), X(s)))\\\nonumber- \ln(\hat{\phi}(1,-1, X(t), X(s))), 
\end{eqnarray}
where $\hat{\phi}(u,v,X(t),X(s))$ is an estimator of the characteristic function:
\begin{eqnarray}
\hat{\phi}(u,v,X(t),X(s))=<\exp\{i(uX(t)+vX(s))\}>.
\end{eqnarray}
In \cite{bib:Rosadi} an efficient methodology is introduced for estimating the codifference 
from a single trajectory of stationary 
process. Namely, if 
$\{x_k, k=1,...,N\}$ is realization of a stationary process $\{X(t)\}$, then the estimator of the characteristic function takes the form:
\begin{eqnarray}
\label{PhiFunction}
\hat{\phi}(u,v,X(t),X(s))=\frac{1}{N}\sum_{k=1}^{N-|t-s|}\exp(i(ux_{k+|t-s|}+vx_{k})).
\end{eqnarray}
For a nonstationary process we are not able to estimate empirical autocodifference by using only a single trajectory, therefore 
the above estimator requires modification. Suppose, we have $M$ trajectories of a nonstationary process $\{X(t)\}$. Let us take 
a sample $\{x^t_k, k=1,...,M\}$ being a realization of a random variable $X(t)$, that is the values of the process $\{X(t)\}$
taken at a fixed time $t$, and a sample $\{x^s_k: k=1,...,M\}$ composed from the values of the process $\{X(t)\}$
taken at a fixed time $s$. By construction, both samples consist of independent identically distributed 
random variables.
%but for each $i=1,2,...,N$ and $j=1,2,...,N$ $y_i$, and $z_j$ are not independent because $X(t)$ and $X(s)$ are not independent. 
Thus, in the nonstationary case the estimator of characteristic function $\phi(u,v,X(t),X(s))$ is defined as:
\begin{eqnarray}
\label{PhiFunction1}
\hat{\phi}(u,v,X(t),X(s))=\frac{1}{M}\sum_{k=1}^M\exp(i(ux^t_k+vx^s_k)).
\end{eqnarray}
%This unpleasant property forces us to simulate in each time steps $s$ and $t$ the one dimensional samples from the distributions $F_{X(s)}(\cdot)$ and $F_{X(t)}(\cdot)$ of the series. Having such samples we are able to apply given above estimator and obtain an estimate of codifference between $X(s)$ and $X(t)$.
%From \cite{bib:Rosadi} (Theorem 1) we infer that $\widehat{CD}$ is (weakly) consistent estimator of $CD$, i.e. 
% it converges in probability to the true value of this parameter. \textcolor{red}{Maybe now is better? 
%For us it is obvious when we have nonstationary process and want to calculate some measure between $X(t)$ and $X(s)$ you need 
%realizations: one from distribution corresponding to $X(t)$ and the second corresponding to $X(s)$ and then on the basis 
%of them we use formula similar to (\ref{PhiFunction}). }

\begin{figure}[thp]
\centering
\includegraphics[height=8.2cm]{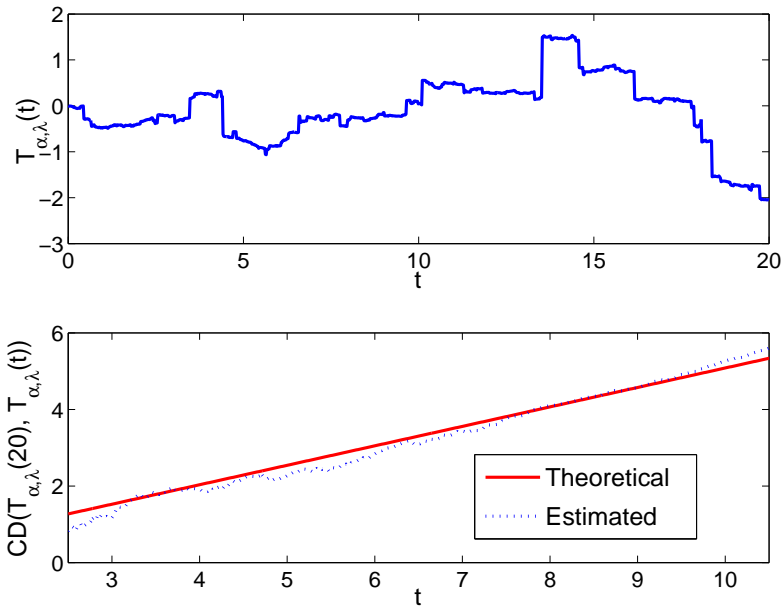}
\caption{Top panel: single trajectory of the tempered stable L\'evy motion $\{T_{\alpha,\lambda}(t)\}$ with parameters $\alpha=0.9, \lambda=0.01$. 
Bottom panel: Estimator of autocodifference $CD(T_{\alpha,\lambda}(t), T_{\alpha,\lambda}(20))$ calculated with $1000$ trajectories, and the 
theoretical value given by Eq.(\ref{tscd}). 
 }
\label{figure1}
\end{figure}

\begin{figure}[thp]
\centering
\includegraphics[height=8.2cm]{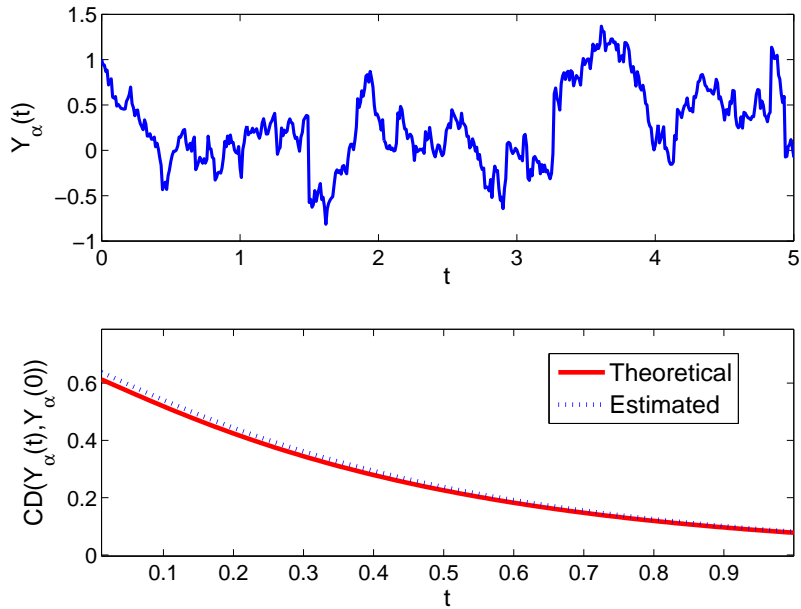}
\caption{The trajectory of stable L\'evy Ornstein-Uhlenbeck process $\{Y_\alpha(t)\}$ with parameters $\alpha=1.6382, \lambda=0.0045$ (top panel),
together with theoretical autocodifference $CD(Y_\alpha(t), Y_\alpha(0))$ and its estimator (bottom panel).
 }
\label{figure2}
\end{figure}

\begin{figure}[thp]
\centering
\includegraphics[height=8.2cm]{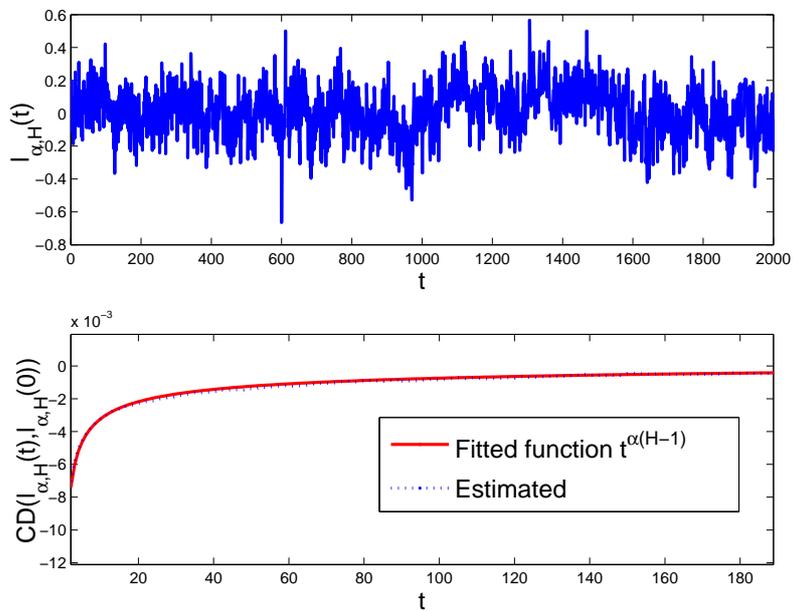}
\caption{The trajectory of fractional L\'evy  noise $l_{\alpha,H}(t)$ with parameters $\alpha=1.95, H=0.8$ (top panel), together with 
the estimator of autocodifference $CD(l_{\alpha,H}(t), l_{\alpha,H}(0))$ and fitted power function $t^{\alpha(H-1)}$ (bottom panel).
 }
\label{figure3}
\end{figure}

In Fig. \ref{figure1} we show a single trajectory of the tempered stable L\'evy motion and compare estimator of the autocodifference 
obtained from Eqs.(\ref{CodifferenceEstimator}) and (\ref{PhiFunction1}) with theoretical value given by Eq.(\ref{tscd}). One observes perfect agreement 
between theoretical and empirical results.
In Fig. \ref{figure2} we present a trajectory of L\'evy stable Ornstein-Uhlenbeck process together with the theoretical 
autocodifference given by Eq.(\ref{cd_ou}) and its estimator obtained from Eqs.(\ref{CodifferenceEstimator}) and (\ref{PhiFunction}). 
In Fig. \ref{figure3} we plot the path of  fractional L\'evy noise and also compare the theoretical autocodifference, Eq.(\ref{fln}), and its estimator. 
In all three cases one observes almost perfect agreement between the empirical and theoretical results. 

%\begin{itemize}
%\item Formula for estimator and explanation how it works
%\item Simulations for finite varince processes (Tempered stable L\'evy process)
%We fixed one variable (i.e. t) and plot codifference as a function of s.
%\item Simulations for infinite varince processes (eg. L\'evy Ornstein-Uhlenbeck process, fractional L\'evy motion)
%\item Plot with simulated sample and empirical codifference together with theoretical one - 3 figures (each figure has two panels: trajectory and codifferences)
%\end{itemize}
\section{Real data analysis}
 \subsection{Plasma data}

\begin{figure}[thp]
\centering
\includegraphics[height=4.2cm]{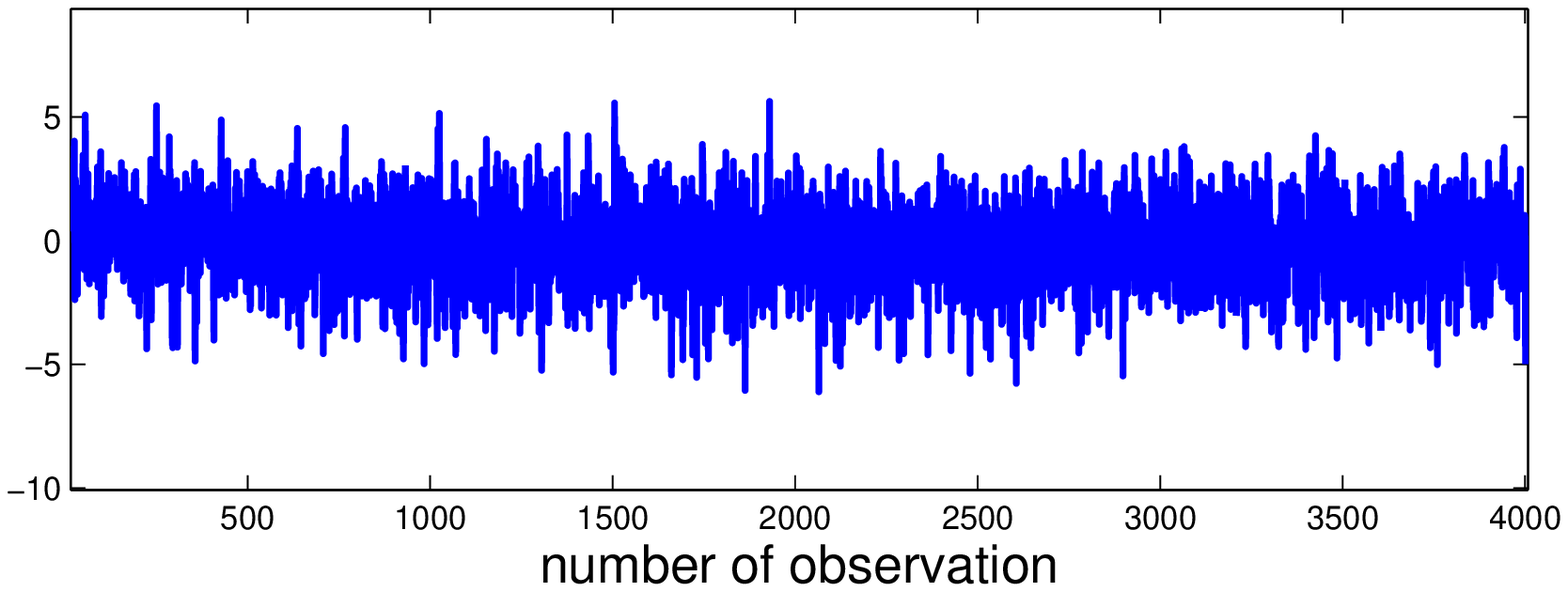}
\includegraphics[height=4.2cm]{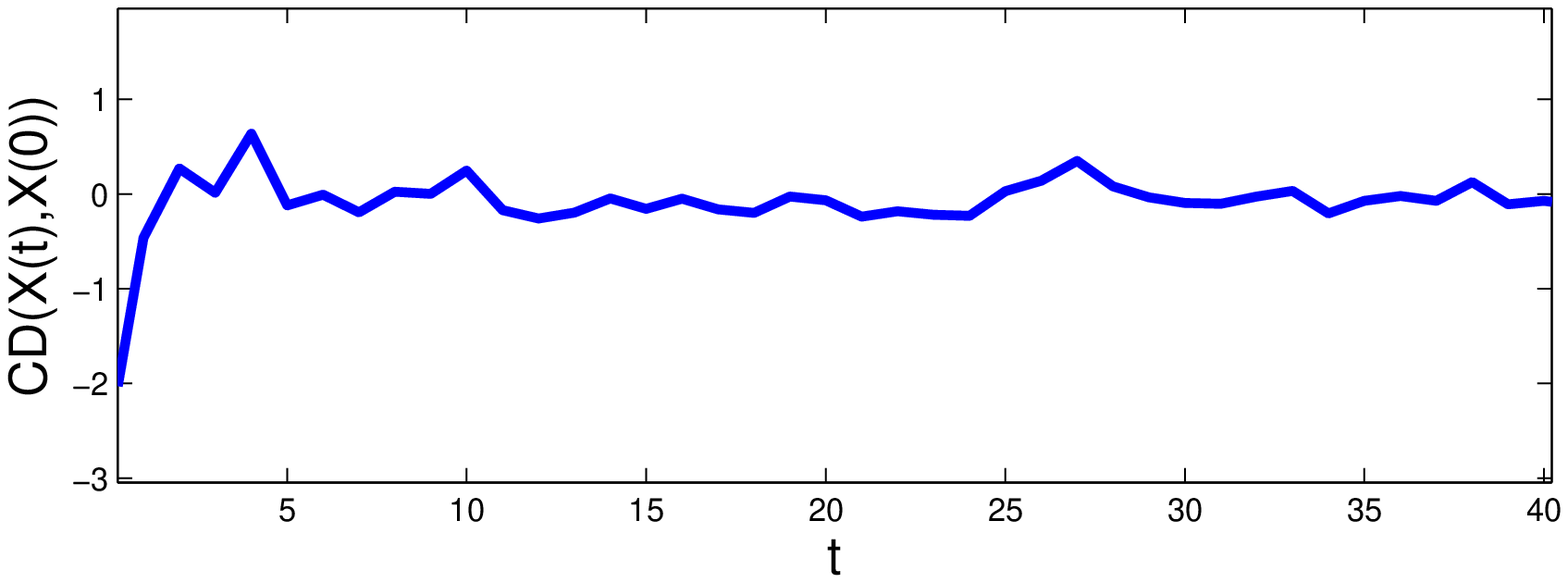}
%\vspace*{-3cm}
\caption{The examined time series of the ion saturation current fluctuations (in
mA) registered in the U-3M torsatron at the small torus radial positions $r = 9.9 cm$ 
(top panel), and the estimator of autocodifference (bottom panel).}
\label{fig_real22}
\end{figure}

In this section we investigate the data obtained in experiment
on the controlled fusion device. The important characteristics of edge plasma turbulence, 
such as fluctuation amplitudes, spectra, and turbulence-induced transport are investigated in the Uragan-3M (U-3M) 
stellarator torsatron by the use of high resolution measurements of density (ion saturation current) and potential
(floating potential) fluctuations with the help of movable
Langmuir probe arrays. We address the reader to \cite{nasza36} for the details of experimental set-up and description
of the data base. Here we present the analysis of the ion saturation current fluctuations (in mA) 
measured at the small torus radial position $r = 9.9 cm$. Similar data were analyzed in \cite{plasma4}. 
In Fig. \ref{fig_real22}  we present the examined time series (top panel) and corresponding estimator of the 
autocodifference (bottom panel).\\
The estimator of autocodifference has non-zero value at $t=0$  and then drops to zero.
Therefore, we infer 
that the data can be considered as a white noise. Moreover, we also observe non-Gaussian  behavior of the underlying series.
Thus, we propose to model the process by using white L\'evy noise. We confirm our assumption 
by using the Anderson-Darling goodness-of-fit test for stable distribution \cite{reg}, which indicates 
that the data come from L\'evy stable distribution (p-value is equal to $0.88$). Then, by using the regression 
method \cite{reg} we estimate the index of stability. As a result we get $\hat{\alpha}=1.95$.

On the basis of autocodifference we can also estimate the scale parameter $\sigma$ of the analyzed series. 
In order to do this we compare the value of the estimator of autocodifference for $t=0$ with the theoretical value 
$-2\sigma^{\alpha}$. As the result we obtain $\hat{\sigma}=1.05$. We then compare the value obtained 
with the estimates of $\sigma$ parameter made by other methods, namely, regression \cite{reg} and McCulloch \cite{mc} methods.
The obtained values are $\hat{\sigma}=1.06$ and $\hat{\sigma}=1.03$, respectively, 
which is in good agreement with the value obtained from the autocodifference.
%In Fig. \ref{fig_real22} we present the empirical codifference of examined plasma data compared with the theoretical one given in formula (\ref{cd_sn}).
\subsection{Financial data}
\begin{figure}[thp]
\centering
\includegraphics[height=4.2cm]{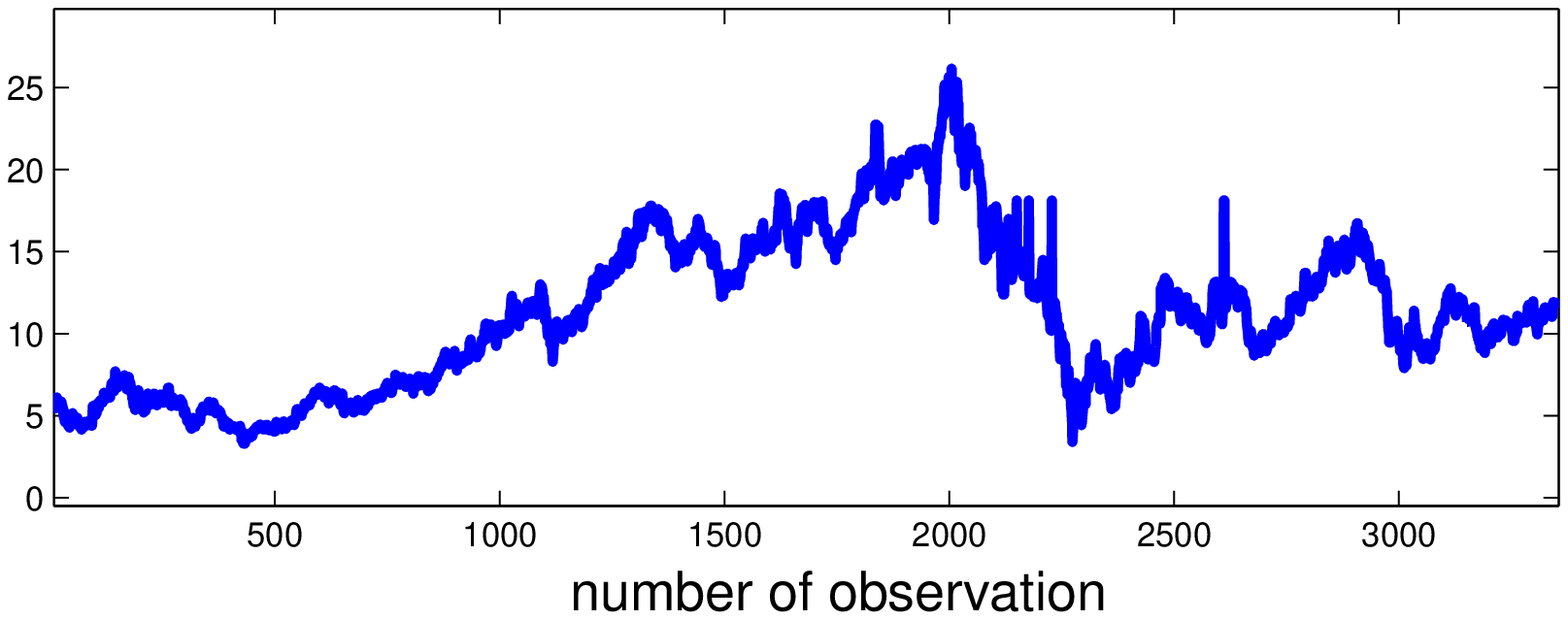}
\includegraphics[height=3.6cm]{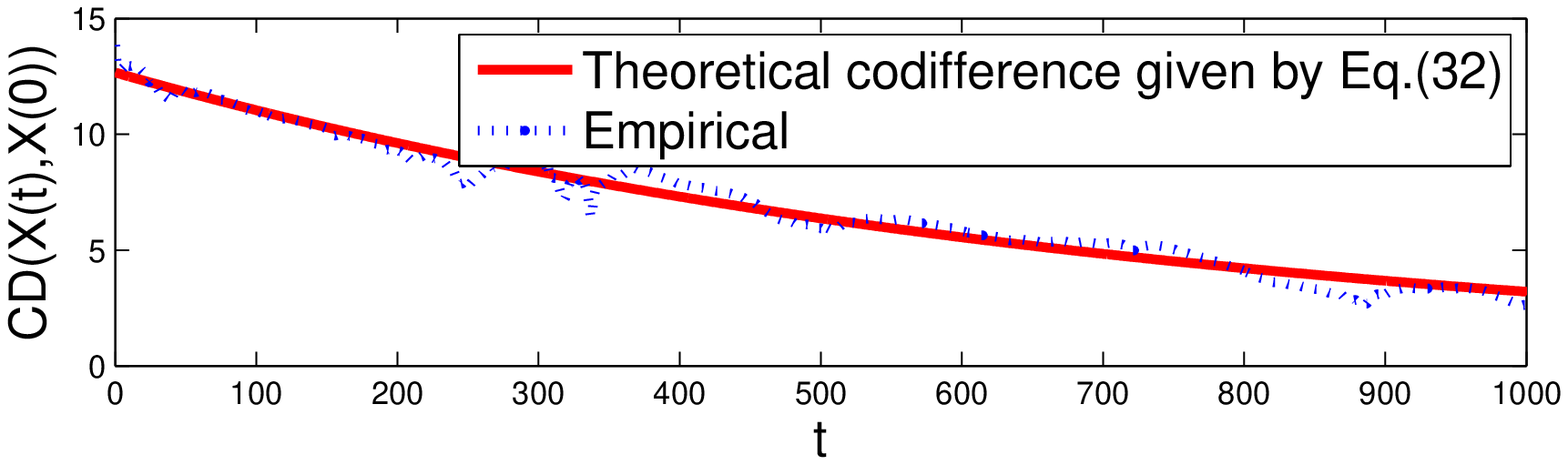}
%\vspace*{-3cm}
\caption{The examined time series that describes closing prices (in USD) of Cosco Pacific Ltd. in the period 04.01.2000-14.01.2013 (top panel), 
and the estimator of autocodifference and theoretical function given by Eq.(\ref{cd_ou}) (bottom panel). }
\label{fig_real1}
\end{figure}

As a second example we analyze time series that describes closing prices of the investment holding company Cosco Pacific Ltd. 
The data are quoted daily in the period 04.01.2000-14.01.2013, \cite{finance}. In Fig. \ref{fig_real1} (top panel) 
we present the examined time series.\\

We recall that the Gaussian Ornstein-Uhlenbeck process originally used by Vasicek for the analysis of interest rates \cite{vas},
was successfully applied to many other data from financial markets \cite{barndorff}. In line with these findings, we intend to
describe the data analyzed by using Ornstein-Uhlenbeck process. However, in contrast to the classical model, significant jumps 
in the examined data are clearly observed,
which may indicate non-Gaussian behavior of the process. We thus suggest the L\'evy Ornstein-Uhlenbeck process 
defined in (\ref{o-u-s}) as a candidate for fitting the data after removing the mean.
%\begin{figure}[thp]
%\centering
%\includegraphics[height=4.2cm]{figfin_cd.eps}
%\caption{The empirical codifference for analyzed financial time series.}
%\label{fig_real2}
%\end{figure}
At the first step of our analysis we estimate the relaxation parameter $\lambda$ that 
enters Eq.(\ref{o-u-s}). Here we employ the Whittle estimation method  described in \cite{klup} (also 
used in \cite{janczura_orzel_wylomanska}), which is based on the sample periodogram of the analyzed time series.  
Once the $\hat{\lambda}$ estimate is found, the residuals of the process can be derived (recall that in our case the residual 
of the L\'evy Ornstein-Uhlenbeck model is a process $l_{\alpha}(t)$ in (\ref{o-u-s})). On the basis of the residual series, 
by using the regression method,  we can estimate the index of stability $\alpha$. 
We also confirmed the stable distribution of the residual series by using the Anderson-Darling goodness of fit test.  
As the result, we obtain $\hat{\lambda}=0.0045$ and $\hat{\alpha}=1.64$. In Fig. \ref{fig_real1} we present 
the estimator of autocodifference along with the theoretical value given by Eq.(\ref{cd_ou}).

Similar to the plasma data analysis, we estimate the scale parameter $\sigma$
from the autocodifference function by using estimated values of $\lambda$ and $\alpha$ in Eq.(\ref{cd_ou}).
As the result we obtain $\hat{\sigma}=0.17$, which is exactly the same value that
the regression and McCulloch methods give for these data.

\section{Conclusions}
In this paper we have examined the codifference, a general measure of interdependence, 
that can be considered 
as an alternative to covariance function. We have indicated the importance of codifference 
especially for processes 
with non-Gaussian distribution, for which the correlation function is not defined. 
For a class
of Gaussian processes the autocodifference is simply reduces to autocovariance with negative sign. 
We then
analyzed this measure for several well-known processes with finite variance and show close similarity between 
autocodifference and autocovariance. We furthermore present the generic examples of the processes, for which
the autocovariance does not exist, therefore the codifference becomes of particular importance. 
After giving a simple 
practical recipe how to estimate the autocodifference from experimental data for both stationary and nonstationary 
processes, we estimated codifference from the surrogate data and analyze two real data sets representing random
fluctuations observed in laboratory plasma and prices on financial market. For both examples we demonstrated 
how the estimated autocodifference can be used as a  tool for recognition a proper stochastic model. 
Also, we have shown 
that on the basis of codifference it is possible to estimate parameters of the fitted model.
Summarizing, we conclude that the codifference serves as a convenient practical tool to
study interdependence for stochastic processes with infinite and finite variances as well. 
\section*{Acknowledgements} The research of Agnieszka Wy{\l}oma{\'n}ska is co-financed by
the National Science Center, Poland, under the contract No. UMO-20127/B/ST8/03031.

\section*{Appendix}
\begin{description}
\item[A1.]{\textit{Tempered stable L\'evy process}}\\
Since the process $\{T(t)\}$ has stationary independent increments, zero mean, and  $<T^2(t)>=2t\alpha(\alpha-1)\lambda^{\alpha-2}$, then for $s<t$
we obtain 
\[cov(T(t),T(s))=<T(t)(T(s)>=<T(s)(T(t)-T(s))>+<T^2(s)>=2s\alpha(\alpha-1)\lambda^{\alpha-2}.\]
Using the same reasoning for $t<s$ we get $cov(T(t),T(s))=2\alpha(\alpha-1)\lambda^{\alpha-2}\min\{s,t\}$.\\
Now let us calculate characteristic function of $T(t)-T(s)$ for $s<t$. We have
\[<\exp\{i(T(t)-T(s))\}>=<\exp\{iT(t-s)\}>=\exp\left\{(t-s)\left[(\lambda+i)^{\alpha}+(\lambda-i)^{\alpha}-2\lambda^{\alpha}\right]\right\}.\]
Moreover,
\[<\exp\{iT(t)\}>=\exp\left\{t\left[(\lambda+i)^{\alpha}+(\lambda-i)^{\alpha}-2\lambda^{\alpha}\right]\right\},\]
\[<\exp\{-iT(s)\}>=\exp\left\{s\left[(\lambda+i)^{\alpha}+(\lambda-i)^{\alpha}-2\lambda^{\alpha}\right]\right\}.\]
As a final result, we obtain:
\[CD((T(t),T(s))=-(t-s)\left[(\lambda+i)^{\alpha}+(\lambda-i)^{\alpha}-2\lambda^{\alpha}\right]+t\left[(\lambda+i)^{\alpha}+(\lambda-i)^{\alpha}-2\lambda^{\alpha}\right]+s\left[(\lambda+i)^{\alpha}+(\lambda-i)^{\alpha}-2\lambda^{\alpha}\right]\]
\[=2s[(\lambda-i)^{\alpha}+(\lambda+i)^{\alpha}-2\lambda^{\alpha}].\]
Therefore, we have
\[CD((T(t),T(s))=2\min\{t,s\}[(\lambda-i)^{\alpha}+(\lambda+i)^{\alpha}-2\lambda^{\alpha}].\]
\item[A2.]{\textit{Laplace motion}}\\
The gamma process $\{G(t)\}$ with parameters $\gamma, \lambda$ is a pure jump L\'evy increasing process with the moments
\begin{equation}
\label{momentGamma}
E(G^n(t))=\lambda^{-n}{\Gamma}(\gamma t+n)/{\Gamma}(\gamma t), n\geq 0,
\end{equation}
and ${\Gamma}(\cdot)$ is a Gamma function. 
If $\Lambda(t)=B(G(t))$, then $<\Lambda(t)>=0$, and we have
\begin{eqnarray*}
cov(\Lambda(t),\Lambda(s))=<\Lambda(t)\Lambda(s)>&=&<(\Lambda(t)-\Lambda(s)+\Lambda(s))\Lambda(s)>=<(\Lambda(t)-\Lambda(s))\Lambda(s)+\Lambda^2(s)>\\
&=&<\Lambda^2(s)>=<B^2(u)|G(s)=u>= <G(s)>=\frac{{\Gamma}(\gamma s+1)}{\lambda{\Gamma}(\gamma s)}.
\end{eqnarray*}
Let us calculate the autocodifference of the process $\{\Lambda(t)\}$. For  $s<t$ we have:
\begin{eqnarray*}
\ln <\exp(i(\Lambda(t)-\Lambda(s))>&=&\ln <\exp(i(\Lambda(t-s))>=\ln <\exp(iB(u))|G(t-s)=u>\\
&=&\ln <\exp(-1/2\Gamma(t-s))>= (-\gamma(t-s))\ln (1+1/(2\lambda)).
\end{eqnarray*}
Thus the autocodifference for $s<t$ is equal to 
\begin{equation*}
\begin{split}
CD(\Lambda(t),\Lambda(s))= -\gamma t\ln (1+1/(2\lambda))-\gamma s\ln (1+1/(2\lambda))+\gamma(t-s)\ln (1+1/(2\lambda))\\
=-2\gamma s\ln (1+1/(2\lambda)).
\end{split}
\end{equation*}

\item[A3.]{\textit{L\'evy Ornstein-Uhlenbeck process}}\\
We note that the random variable $Y_{\alpha}(t)$ given by (\ref{rozw11}) for $\lambda>0$ is S$\alpha$S with 
 $\sigma^{\alpha}=\frac{\sigma^{\alpha}}{\alpha\lambda}$. This fact  follows directly 
from the Propositions 3.4.1  and 3.5.2 in \cite{samorodnitsky}.  To obtain the  formula 
for the autocodifference, we use the relation 
between the scale parameters and codifference given in (\ref{cod1}). Then we obtain:
\begin{eqnarray*}
\sigma_{Y_{\alpha}(t),Y_{\alpha}(t)}^{\alpha}=\sigma_{Y_{\alpha}(s),Y_{\alpha}(s)}^{\alpha}=\frac{\sigma^{\alpha}}{\lambda\alpha}.
\end{eqnarray*}
From Eq.(\ref{rozw11}) for $s<t$ we get:
\[Y_{\alpha}(t)-Y_{\alpha}(s)=\int_{-\infty}^{\infty}f(t-x)-f(s-x)dL_{\alpha}(x),\]
where $f(t-x)=e^{-\lambda(t-x)}1_{\{x<t\}}$. Using Proposition 3.5.2 in  \cite{samorodnitsky} 
we obtain the scale parameter of random variable $Y_{\alpha}(t)-Y_{\alpha}(s)$ for $s<t$:
\[\sigma_{Y_{\alpha}(t)-Y_{\alpha}(s)}^{\alpha}=\sigma^{\alpha}\int_{-\infty}^{\infty}|f(t-x)-f(s-x)|^{\alpha}dx.\]
Therefore we have:
\begin{eqnarray*}\sigma_{Y_{\alpha}(t)-Y_{\sigma}(s)}^{\alpha}&=&\sigma^{\alpha}\int_{-\infty}^s|f(t-x)-f(s-x)|^{\alpha}dx+\sigma^{\alpha}\int_{s}^t|f(t-x)|^{\alpha}dx\\
&=&\sigma^{\alpha}\int_{-\infty}^s|e^{-\lambda(t-x)}-e^{-\lambda(s-x)}|^{\alpha}dx+\sigma^{\alpha}\int_{s}^te^{-\lambda\alpha(t-x)}dx\\
&=&\sigma^{\alpha}|e^{-\lambda t}-e^{-\lambda s}|^{\alpha}\int_{-\infty}^se^{\lambda\alpha x}dx+\sigma^{\alpha}\frac{1-e^{-\lambda\alpha(t- s)}}{\lambda\alpha}
\\&=&\sigma^{\alpha}\frac{|1-e^{-\lambda(t-s)}|^{\alpha}+1-e^{-\lambda\alpha(t- s)}}{\lambda\alpha}.
\end{eqnarray*}
Then we obtain: 
\begin{eqnarray*}
CD(Y_{\alpha}(t),Y_{\alpha}(s))&=&-\sigma^{\alpha}\frac{1+e^{-\lambda\alpha(t- s)}-|1-e^{-\lambda(t-s)}|^{\alpha}}{\lambda\alpha}.
\end{eqnarray*}

\item[A4.]{\textit{Fractional L\'evy motion}}\\
From Proposition 3.4 in \cite{maej} we infer that for the stable integrals 
$\int_{S}f(x)dL_{\alpha}(x)$, where $\{L_{\alpha}(t)\}$ is a symmetric L\'evy motion with $0<\alpha<2$ and $S\subset R$, the following holds:
\begin{eqnarray}\label{maej1}
<\exp\left\{i\theta \int_{S}f(x)dL_{\alpha}(x)\right\}>=\exp\left\{-|\theta|^{\alpha}\int_{S}|f(x)|^{\alpha}dx\right\}
\end{eqnarray}
if $\int_{S}|f(x)|^{\alpha}dx<\infty$.\\
Using formula (\ref{maej1}) and the fact that fractional L\'evy motion $\{L_{\alpha,H}(t)\}$ is self similar and has stationary increments,
 for $t<s$ we obtain:
\[CD(L_{\alpha,H}(t),L_{\alpha,H}(s))=\ln(<exp\{it^{H}L_{\alpha,H}(1)\}>)+\ln(<exp\{-is^{H}L_{\alpha,H}(1)\}>)-\ln(<exp\{i(t-s)^{H}L_{\alpha,H}(1))\}>).\]
Therefore, we have:
\[CD(L_{\alpha,H}(t),L_{\alpha,H}(s))=\left(|t-s|^{\alpha H}-|t|^{\alpha H}-|s|^{\alpha H}\right)k(H,\alpha),\]
where 
\[k(H,\alpha)=\int_{-\infty}^{\infty}\left| \max(1-u,0)^{H-1/\alpha}-\max(-u,0)^{H-1/\alpha}\right|^{\alpha}du.\]
Now we show that the above integral converges. Indeed, we have
\begin{eqnarray*}k(H,\alpha)&=&\int_{-\infty}^0\left| (1-u)^{H-1/\alpha}-(-u)^{H-1/\alpha}\right|^{\alpha}du+\int_{0}^1|1-u|^{H\alpha-1}du=\int_{0}^{\infty}\left| (1+u)^{H-1/\alpha}-u^{H-1/\alpha}\right|^{\alpha}du+\frac{1}{H\alpha}\\
&=&\int_{0}^{1}\left| (1+u)^{H-1/\alpha}-u^{H-1/\alpha}\right|^{\alpha}du+\int_{1}^{\infty}\left| (1+u)^{H-1/\alpha}-u^{H-1/\alpha}\right|^{\alpha}du+\frac{1}{H\alpha}.\end{eqnarray*}
The first integral converges, because
\[\int_{0}^{1}\left| (1+u)^{H-1/\alpha}-u^{H-1/\alpha}\right|^{\alpha}du\leq \int_{0}^{1} (1+u)^{H\alpha-1}du=\frac{2^{H\alpha}}{H\alpha}.\]
Note that
\[(1+u)^{H-1/\alpha}-u^{H-1/\alpha}=(H-1/\alpha)\int_{0}^1(x+u)^{H-1/\alpha-1}du.\]
Therefore, for $H\neq 1/\alpha$ we have
\[\left|(1+u)^{H-1/\alpha}-u^{H-1/\alpha}\right|^{\alpha}=|(H-1/\alpha)|^{\alpha}\left|\int_{0}^1(x+u)^{H-1/\alpha-1}dx\right|^{\alpha}\leq |(H-1/\alpha)|^{\alpha}|1+u|^{\alpha(H-1)-1},\]
which gives for $H<1$
\[\int_{1}^{\infty}\left| (1+u)^{H-1/\alpha}-u^{H-1/\alpha}\right|^{\alpha}du \leq \int_{1}^{\infty}|(H-1/\alpha)|^{\alpha}|1+u|^{\alpha(H-1)-1}du=\frac{|(H-1/\alpha)|^{\alpha}}{\alpha(1-H)}.\]
\item[A5.]{\textit{Supediffusion continuous time random walk-like process}}\\
Let us first calculate the characteristic function of $Z(t)$. We have:
\[ \Phi_{Z(t)}(k)=<\exp\{-\frac{L_{\alpha}(t)k^2}{2}\}>.\]
Now taking the formula of the Laplace transform of the process $\{Z(t)\}$ given in (\ref{lt}), we obtain:
\[ \Phi_{Z(t)}(k)=\exp\{-tk^2\sigma^{\alpha}2^{-\alpha}\}.\]
Since  $Z(t)-Z(s)\sim Z(t-s)$ for $t>s$ ($"\sim"$ means equality in distribution), then   we obtain:
\[CD(Z(t),Z(s))=-t\sigma^{\alpha}2^{-\alpha}-s\sigma^{\alpha}2^{-\alpha}+(t-s)\sigma^{\alpha}2^{-\alpha}=-\sigma^{\alpha}2^{-\alpha+1}s=-\sigma^{\alpha}2^{-\alpha+1}\min\{t,s\}.\]

\end{description}

\end{document}